\documentclass[floatfix,%
 reprint,
superscriptaddress,
 amsmath,amssymb,
 aps,
pra,
]{revtex4-2}

\usepackage{amsmath,amssymb,amsfonts}
\usepackage{algorithmic}
\usepackage{graphicx}
\usepackage{svg}
\usepackage{textcomp}
\usepackage{xcolor}
\usepackage{hyperref}
\usepackage{physics}
\usepackage{diagbox}
\usepackage{bm}
\usepackage{array}
\usepackage[normalem]{ulem} 
\usepackage{tikz}
\usetikzlibrary{decorations.pathreplacing}
\usepackage{slashbox}
\usepackage{graphicx}
\usepackage{dcolumn}
\usepackage{ragged2e}
\usepackage{placeins}
\usetikzlibrary{quantikz2, arrows.meta, positioning, calc}

\usepackage[font=footnotesize,justification=raggedright,singlelinecheck=false]{caption}
\usepackage{subcaption}
\newcommand{\WF}{\psi(\bm{\theta})}

\newcommand{\BigO}{\mathcal{O}}
\newcommand{\ER}{Erd\"{o}s-R\'{e}nyi }
\newcommand{\Mod}{\ \mathrm{mod}\ }

\newcommand{\sidecaption}[1]
{\raisebox{\abovecaptionskip}{\begin{subfigure}[t]{1.6em}
			\caption[singlelinecheck=off]{}
			\label{#1}
	\end{subfigure}}\ignorespaces}

\def\BibTeX{{\rm B\kern-.05em{\sc i\kern-.025em b}\kern-.08em
    T\kern-.1667em\lower.7ex\hbox{E}\kern-.125emX}}
\usepackage{indentfirst}

\begin{document}
\title{Light Cone Cancellation for Variational Quantum Eigensolver in Solving Noisy Max-Cut}

\author{Xinwei Lee}
\email[]{xwlee@smu.edu.sg}
\affiliation{School of Computing and Information Systems, Singapore Management University}
\author{Xinjian Yan}
\email[]{yanxinjian@cavelab.cs.tsukuba.ac.jp}
\affiliation{Graduate School of Science and Technology, University of Tsukuba}
\author{Ningyi Xie}
\affiliation{Graduate School of Science and Technology, University of Tsukuba}
\author{Yoshiyuki Saito}
\author{Leo Kurosawa}
\affiliation{Graduate School of Computer Science and Engineering, University of Aizu}
\author{Nobuyoshi Asai}
\affiliation{School of Computer Science and Engineering, University of Aizu}
\author{Dongsheng Cai}
\affiliation{Faculty of Engineering, Information and Systems, University of Tsukuba}
\author{Hoong Chuin LAU}
\affiliation{School of Computing and Information Systems, Singapore Management University}

\begin{abstract}
Variational Quantum Eigensolver (VQE) is a quantum-classical hybrid algorithm used to estimate the ground energy of a given Hamiltonian.
It consists of a parameterized quantum circuit, which the parameters are optimized using a classical optimizer.
With the increasing need in solving large-scale problems in real-world applications, solving those large problems with fewer qubits and fewer gates becomes essential,
so that we reduce the simulation difficulty and mitigate the effect of noise in real quantum hardware.
In this study, we applied the Light Cone Cancellation (LCC) method to reduce the number of qubits and gates required in a two-local ansatz.
LCC removes redundant gates that are not required in the calculation of the expectation value for a local observable.
This leads to two consequences: 1) the quantum circuit used to create the trial wavefunction of VQE can be broken down into multiple quantum subcircuits with fewer qubits,
enabling large-scale problems to be solved without actually simulating the entire circuit;
and 2) reduced number of quantum gates in the circuit leads to the noise mitigation in quantum hardware.
The main purpose of this work is to demonstrate the effectiveness of this method (called the LCC-VQE) in mitigating the device noise when solving the Max-Cut problem up to 100 qubits,
using simulations on small (7-qubit and 27-qubit) fake noisy backends.
Employing a single-layer two-local ansatz circuit architecture, 
the reuslts show that LCC-VQE yields higher approximation ratios than those cases without LCC, implying that the effect of noise is mitigated when LCC is applied.
An analysis of more than one layer of two-local ansatz is also performed, but empirical results show that the single-layer ansatz still performs the best among them.
We also compare LCC-VQE under noiseless conditions with the Goemans-Williamson algorithm. 
\end{abstract}

\maketitle

\section{Introduction}
Many combinatorial optimization problems (COP) are considered to be difficult to address using traditional computational approaches. COPs aim to find the optimal 
combination of variables that minimizes (or maximizes) a given objective function, while simultaneously satisfying a set of constraints. Recent years, people focus on 
using quantum-classical hybrid methods, known as the variational quantum algorithms (VQA)~\cite{cerezo2021variational} to heuristically solve COPs. The quantum 
approximate optimization algorithm (QAOA)~\cite{farhi2014quantum} is one of the VQAs that is intensively explored due to its predictable patterns in the variational 
parameters~\cite{leo2020,Cook2020TheQA,bilinear}, and also its relation with quantum annealing~\cite{farhi:qaa,tqa2021}.
Another VQA, the variational quantum eigensolver 
(VQE)~\cite{peruzzo2014variational}, is also capable of solving COPs, although it is better known for its application to quantum chemistry. Unlike QAOA which has a 
problem-dependent ansatz, the structure of the ansatz in VQE is static and does not depend on the problem solved. Moreover, the VQE ansatz offers a greater degree of 
freedom in the sense that it has greater expressibility~\cite{expressibility} and more number of variational parameters compared to the QAOA ansatz.
In our previous work~\cite{weko_233690_1}, we have shown that VQE generally has better performance than QAOA and the Multi-angle QAOA~\cite{herrman2022multi} in solving the Max-Cut problem
using noiseless simulations.

The VQAs have shown the potential quantum advantage on Noisy Intermediate Scale Quantum (NISQ)~\cite{Preskill_2018} devices. However, an increase in the number of qubits often leads to higher error rates when building actual quantum hardware. Although recent advancements have prominently featured quantum error correction algorithms, those involve intricate designs that must effectively address the inherent noise and decoherence in quantum environments. It appears that reducing the number of qubits and gates is a more feasible and efficient approach while maintaining the accuracy of algorithms. 

In this paper, we apply a method known as Light Cone Cancellation (LCC)~\cite{Lowe_2021,leone2024practical} to solve the Max-Cut problem using VQE.
When computing the expectation function of variational circuits, there are many redundant operators that need not be included in the computations. 
LCC exploits the preliminary knowledge of which operators are redundant, so that we do not include them in the calculation at the first place.
LCC is widely applied for QAOA~\cite{short-qaoa}, and also inspired applications in tensor network~\cite{Huang2021,tensor-lykov1,tensor-lykov2,Vidal_2008,lc-tn-te,var-power-tn}
and quantum machine learning~\cite{lc-qml}.

The primary contribution of our work is that we demonstrate the effectiveness of LCC in reducing the number of qubits required for the simulation of quantum circuits,
and also noise mitigation caused by the reduction in the number of gates in the circuit simulation.
We simulate using Qiskit fake noisy backends with 7 qubits and 27 qubits.
The demonstration results show that, compared with original VQE, the implementation of LCC achieves better performance.
Additionally, to further quantify the performance of LCC-VQE under a noiseless condition, we benchmark it against the Goemans–Williamson (GW) algorithm~\cite{goemans1995improved} on 100-vertex graph instances.
The simulation results indicate that LCC-VQE achieves better performance compared to the GW algorithm on denser graphs.

The rest of the paper is structured as follows. Section II provides background information and details the construction of the LCC architecture.
The detailed results of comparative simulations under both noisy and noiseless conditions are articulated in Section III. Section IV contains the concluding remarks of this study.

\section{Background}
\subsection{Max-Cut}
Max-Cut is a fundamental and widely studied NP-hard combinatorial optimization problem in the field of graph theory~\cite{karp2009reducibility}. The primary objective of Max-Cut is to partition the nodes of an undirected graph into two disjoint subsets such that the number of edges connecting the two subsets is maximized. This problem is relevant in various fields, including network design, VLSI layout, community detection, and social network analysis~\cite{jagannath2018max}.

Consider an $n$-node unweighted, undirected graph $G=(V,E)$, where $V$ represents the set of the nodes and $E$ represents the set of the edges of graph $G$. A cut is defined as a partition of the original set $V$ into two subsets. The cost function $C(\mathbf{x})$ to be maximized is the sum of the edges connecting points in the two different subsets, 
which can be expressed as 
\begin{equation}
    C(\mathbf{x}) = \sum_{(i,j)\in E}(x_{i}\oplus x_{j}),
    \label{eqn:cost}
\end{equation}
where $\mathbf{x} = (x_1,x_2,\ldots, x_n)$ and $x_i$ $\in\{0,1\}$ represent the binary variable of node $i$, and $n = |V|$ is the number of nodes in the graph.
The symbol $\oplus$ denotes the XOR (exclusive-OR) operation. 
We want to find the combinations of $\mathbf{x}$ such that the cost function is maximized, i.e. the number of edges cut is maximum.
A brute-force approach on a classical computer would require $\BigO(2^n)$ time to solve this problem.

In the quantum realm, the cost function in Eq.~(\ref{eqn:cost}) can be formulated as the cost Hamiltonian $H_C$, whose expectation value is to be maximized:
\begin{equation}
    H_{C}=\frac{1}{2} \sum_{(i,j)\in E} (I-Z_{i}Z_{j} ),
    \label{eqn:hc}
\end{equation}
where $I$ is the identity matrix of size $2^n\times 2^n$, and $Z_{i}$ represents Pauli-$Z$ observable on qubit $i$.
One node in the graph is mapped to one qubit in the quantum circuit, so it requires $n$ qubits to encode the solution of Max-Cut.

\begin{figure}
    \centering
    \includegraphics[width=0.9\linewidth]{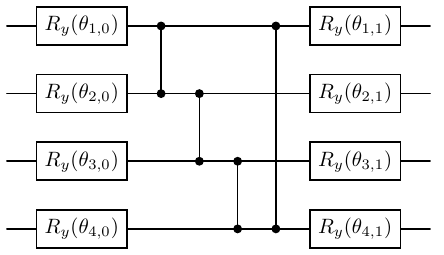}
    \caption{\justifying 
    Two-local ansatz used in the simulations in our work. The architecture of a single layer of $R_y$ gates with circular CZ entanglement is used. The figure shows a 4-qubit example. The parameters $\theta_{k, m}$ are based on the notation specified in Eq.~(\ref{eqn:theta-mat}).}
    \label{fig:twolocal}
\end{figure}

\begin{figure*}[t]
    \centering
    \begin{subfigure}[t]{\textwidth}
        \centering
        \caption*{(a)}
        \vspace{1ex}
        \includegraphics[width=\textwidth]{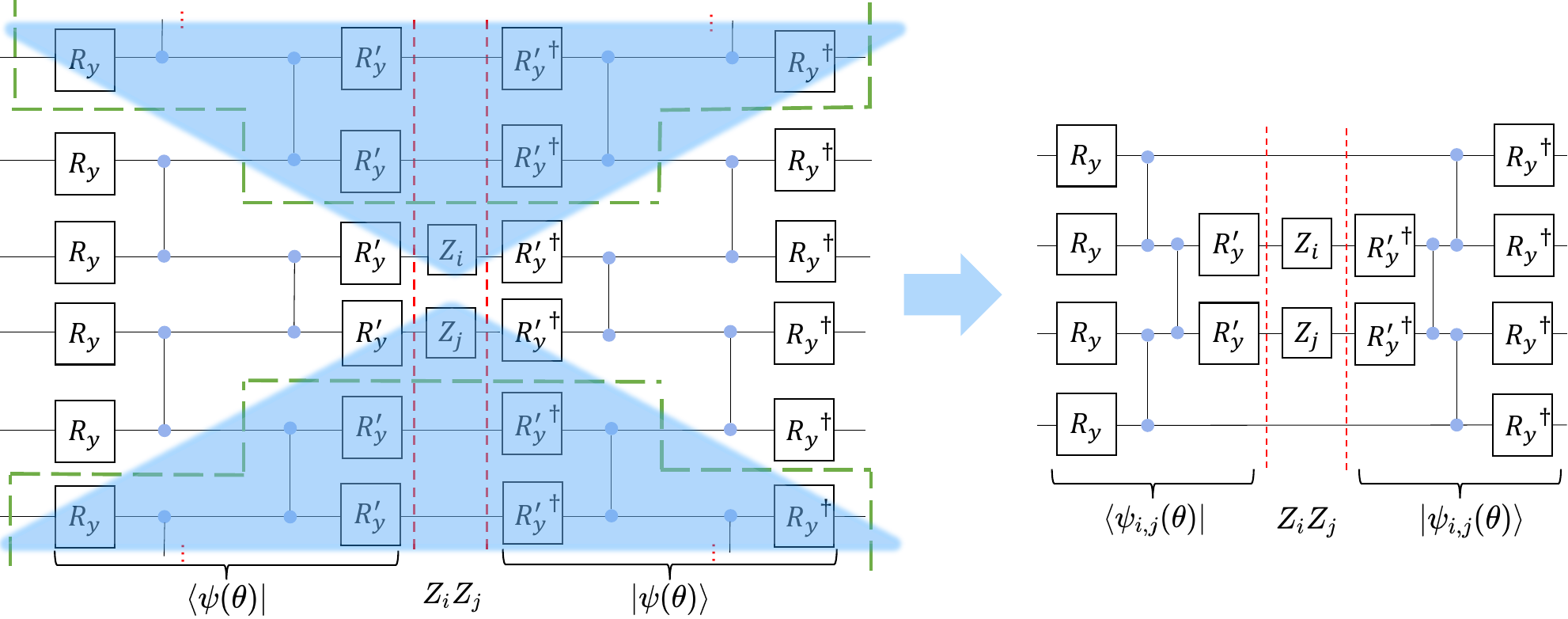}
    \end{subfigure}
    \begin{subfigure}[t]{0.32\textwidth}
		\centering
        \caption*{(b)}
        \vspace{1ex}
		\raisebox{0.3\height}{\includegraphics[width=\textwidth]{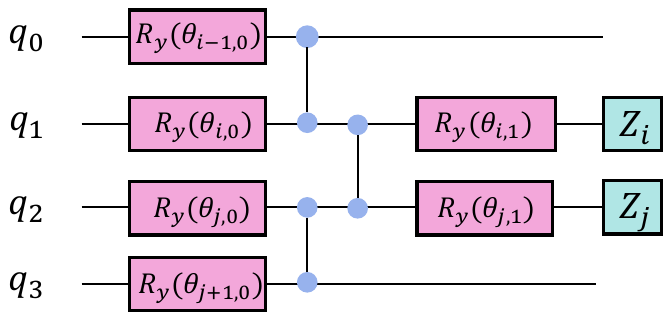}}
	\end{subfigure}
    \hfill
    \begin{subfigure}[t]{0.32\textwidth}
		\centering
        \caption*{(c)}
        \vspace{1ex}
		\raisebox{0.15\height}{\includegraphics[width=\textwidth]{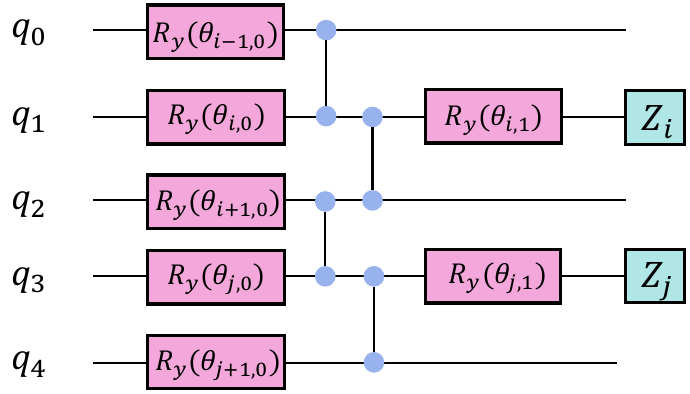}}
	\end{subfigure}
    \hfill
    \begin{subfigure}[t]{0.32\textwidth}
		\centering
        \caption*{(d)}
        \vspace{1ex}
		\includegraphics[width=\textwidth]{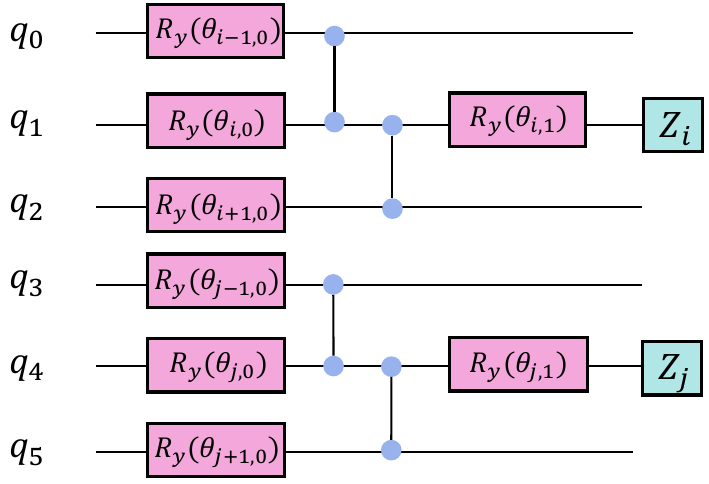}
	\end{subfigure}
    \caption{\justifying
    (a) The light cone cancellation (LCC) in a single layer two-local ansatz. The expectation function $\expval{Z_iZ_j}{\WF}$ is %
    visualized as a quantum circuit on the left figure. $R_y$'s are the single-qubit gates and $C_z$'s are the two-qubit gates. %
    The gates on the left of the red dashed line show $\bra{\WF}$ and the gates on the right show $\ket{\WF}$. %
    The blue shaded regions show the redundant gates that can be cancelled during the calculation of the expectation. %
    The figure on the right shows the resulting circuit after cancellation. (b), (c) and (d) show the possible resulting circuit for the Max-cut Hamiltonian, %
    depending on the indices $i$ and $j$ (positions of the Pauli-Z operators). (b) When $Z_i$ and $Z_j$ are adjacent; (c) when $Z_i$ and $Z_j$ are one qubit apart; %
    and (d) when $Z_i$ and $Z_j$ are two or more qubits apart.}
    \label{fig:lcc}
\end{figure*}

\subsection{Variational Quantum Eigensolver}
The variational quantum eigensolver (VQE) is initially developed for calculating the minimum energy states of molecules.
When re-formulated to address the Max-Cut problem, the expectation value of the cost Hamiltonian $H_C$ over a trial state $\ket{\WF}$ is defined as
\begin{equation}
    \mathcal{E}(\bm{\theta})=\expval{H_C}{\WF}.
    \label{eqn:e-theta}
\end{equation}

The objective is to maximize $\mathcal{E}(\bm{\theta})$, which is equivalent to minimizing $-\mathcal{E}(\bm{\theta})$, using a classical optimizer.
$\bm{\theta}$ is the collection of variational parameters for the VQE ansatz circuit. 
In this paper, we employ a two-local circuit with $R_y(\theta)$ single-qubit rotation gates and CZ (controlled-Z) circular entanglement, with only a single layer. 
Fig.~\ref{fig:twolocal} shows an example of a 4-qubit ansatz circuit.
The circular entanglement has CZ gates between adjacent qubits, and also between the first and the last qubit.
Instead of using a flattened array $\bm{\theta}$, we use a matrix $\Theta$ to represent the variational parameters, aligning them with their geometrical positions in the circuit:
\begin{equation}
    \Theta = 
    \begin{bmatrix}
        \theta_{1,0} & \theta_{1,1} \\
        \theta_{2,0} & \theta_{2,1} \\
        \vdots & \vdots \\
        \theta_{n,0} & \theta_{n,1} 
    \end{bmatrix}.
    \label{eqn:theta-mat}
\end{equation}
It is important to note that $\Theta$ is simply the re-shaped version of $\bm{\theta}$, and the notation $\theta_{k, m}$ denotes the parameters on the $k$-th qubit
in the original circuit. $\theta_{k, 0}$ corresponds to the rotation angle of the initial $R_y$ gate, whereas $\theta_{k, 1}$ corresponds to the rotation angle of the $R_y$ gate applied after the circular CZ entanglement.
Also, due to the periodicity of the $R_y$ rotation gate, the range of the values for $\theta_{k, m}$ are restricted to $[0, 2\pi)$.

Numerous metrics are available for assessing the performance of VQE. Within the scope of unconstrained COP, our focus is on studying the approximation ratio (AR). This metric compares the expected solution obtained through VQE with the optimal solution, essentially measuring how close VQE comes to the best possible outcome. It is defined as
\begin{equation}
    \text{AR} = \frac{\mathcal{E}(\bm{\theta}^*)}{\text{MaxCut}(G)},
\end{equation}
where $\bm{\theta}^*$ is the quasi-optimal parameters returned by the optimizer, $\mathcal{E}(\bm{\theta}^*)$ is its corresponding expectation, and $\text{MaxCut}(G)$ is the exact solution of graph $G$. The closer the value of AR is to 1, the closer it is to the true solution of Max-Cut. 

The performance of VQE and QAOA in addressing Max-Cut problems under noiseless conditions is evaluated in previous research~\cite{weko_233690_1}.
When both algorithms are initialized with the same number of parameters, our findings indicate that VQE outperforms QAOA under random initialization.
Furthermore, the comparison with Multi-angle quantum approximate optimization algorithm (ma-QAOA) reveals that VQE also achieves superior performance on different undirected graphs.

\subsection{Light Cone Cancellation}
Light cone cancellation (LCC) is a method that utilizes the intrinsic property of the expectation function, so that the redundant unitaries in the 
expectation function are not included in its computation in the first place~\cite{benedetti2021hardware}.
The LCC property was originally used in QAOA to reduce the problem graphs to their constituent subgraphs, hence simplifying the problem to be
solved~\cite{brandao2018fixed,leo2022sk,fixed-angle-conjec}.

The following formally describes LCC for the circular-entangled two-local circuit using mathematical derivation.
We start by substituting Eq.~(\ref{eqn:hc}) into the expectation function Eq.~(\ref{eqn:e-theta}), the Max-Cut expectation can be rewritten as
\begin{equation}
     \mathcal{E}(\bm{\theta}) = \frac{|E|}{2} - \frac{1}{2} \sum_{(i,j)\in E} \bra{\psi\bm{(\theta)}} Z_i Z_j \ket{\psi(\bm{\theta)}}.
    \label{eqn:e-theta_1}
\end{equation}
The trial wavefunction (or ansatz) on both sides, $\bra{\WF}$ and $\ket{\WF}$, can be partially cancelled out.
This is because some of the unitary gates used to prepare $\ket{\WF}$ commute through the central local observables $Z_iZ_j$. 
Depending on $i$ and $j$, the state $\ket{\psi(\bm{\theta})}$ prepared by the original full circuit of $L$ alternating layers becomes $\ket{\psi_{i,j}(\bm{\theta})}$, which can be
prepared by subcircuits using less qubits:
\begin{equation}
    \ket{\psi_{i,j}(\bm{\theta})} = U_1 U_{2} \cdots U_L \prod_{s \in \mathcal{S}_{i,j}^{(L)}} R_y(\theta_{s,0})
    \ket{0}^{\otimes n_q},
    \label{eqn:state_psi}
\end{equation}
The operator $U_m$ is the unitary in the $(L-m)$-th layer (larger $m$ means further from the center of the light cone) of the subcircuit and is given by:
\begin{equation}
    U_m = \prod_{s \in \mathcal{S}_{i,j}^{(m)}} R_y(\theta_{s,m}) \prod_{(s,s') \in \mathcal{T}_{i,j}^{(m)}} CZ_{s,s'},
    \label{eqn:um}
\end{equation}
where $R_y(\theta_{s,m})$ denotes a single-qubit $R_y$ gate acting on qubit $s$ at layer $m$, and $CZ_{s,s'}$ denotes a $CZ$ gate acting on qubit $s$ and $s'$.
For all $m\in\{1,2,...,L\}$, we respectively define the sets $\mathcal{S}_i^{(m)}$ and $\mathcal{T}_i^{(m)}$ for the $R_y$ and $CZ$ gates:
\begin{align}
    \mathcal{S}_i^{(m)} & := \{ s\in\{1, 2, \dots, n\} \mid (i - m \leq s \leq i + m)\Mod n \} \\
    \mathcal{T}_i^{(m)} & := \{(s, s') | s \in \mathcal{S}^{(m)}_i, s' = (s+1) \Mod n\}.
\end{align}
These sets contain the qubit indices that span $l$ neighboring qubits of qubit $i$ (because of the observable $Z_i$ on qubit $i$).
Since there are two observables, $Z_i$ and $Z_j$, the qubits spanned by these observables are the union of the two sets, hence we define:
\begin{align}
    \mathcal{S}_{i,j}^{(m)} & := \mathcal{S}^{(m)}_i \cup \mathcal{S}^{(m)}_j\\
    \mathcal{T}_{i,j}^{(m)} & := \mathcal{T}^{(m)}_i \cup \mathcal{T}^{(m)}_j
    \label{eqn:union-set}
\end{align}
to represent the union sets in the subscript of the operators in Eq.~(\ref{eqn:um}). The total number of qubits required for each subcircuit is 
\begin{equation}
    n_q' = \left| \mathcal{S}_{i,j}^{(L)} \right|,
\end{equation}
where $|\cdot|$ is the set cardinality.
Thus, the number of qubits required for each circuit is $\mathcal{O}(L)$.
It is important to know that $n_q'\leq n$, in which LCC leads to a reduction in the number of qubits.

Consequently, the states $\ket{\psi_{i,j}(\bm{\theta})}$ prepared by the subcircuit is used to calculate the expectation $\mathcal{E}(\bm{\theta})$
instead of using the full circuit state $\ket{\psi(\bm{\theta})}$:
\begin{equation}
     \mathcal{E}_\text{LCC}(\bm{\theta}) = \frac{|E|}{2} - \frac{1}{2} \sum_{(i,j)\in E} \bra{\psi_{i,j}(\bm{\theta)}} Z_i Z_j \ket{\psi_{i,j}(\bm{\theta)}}.
    \label{eqn:e-theta_2}
\end{equation}
Since LCC is just cancelling the redundant operators, the expectation computed using LCC is technically the same as the original expectation:
\begin{equation}
    \mathcal{E}(\bm{\theta}) \equiv \mathcal{E}_\text{LCC}(\bm{\theta}).
    \label{eqn:lcc-equals-no-lcc}
\end{equation}

Fig.~\ref{fig:lcc}(a) shows an example of the LCC of a two-local circuit used to prepare the trial wavefunction for VQE. 
The figure visualizes the expectation function Eq.~(\ref{eqn:e-theta}) as a quantum circuit. The circuit on the left of the red dashed line shows the term $\bra{\WF}$, 
and the circuit on the right of the dashed line shows $\ket{\WF}$, with the central observable $Z_iZ_j$ (between the red dashed lines) acting on qubit $i$ and $j$.
Since $\ket{\WF}$ is just the conjugate transpose of $\bra{\WF}$, they are the counterpart of each other in the circuit.
The blue shaded regions indicate the gates that are not related to qubits $i$ and $j$ can commute through the center are cancelled.
The result of this cancellation is shown on the right side of Fig.~\ref{fig:lcc}(a), which has reduced the number of qubits and gates.

The Max-Cut Hamiltonian in Eq.~(\ref{eqn:hc}) has a two-local $Z$ observable on every term.
After LCC on the two-local ansatz that we considered (single layer $R_y$ and circular CZ entanglement),
we can get 3 different types of subcircuits as shown in Fig.~\ref{fig:lcc}(b), (c) and (d),
which requires 4-, 5-, and 6-qubit quantum circuits, respectively.
Circular entanglement means the adjacent qubits, as well as the first and the last qubits, are entangled. 
Note that one subcircuit corresponds to the expectation of a local term in the Hamiltonian, so the simulation of the subcircuits can be done separately.
Also, the subcircuit in Fig.~\ref{fig:lcc}(d) can be further divided into two separate circuits as the first 3 qubits and the last 3 qubits are not entangled. 
Thus, we only require a maximum of 5 qubits to simulate the expectation of the entire Max-Cut Hamiltonian, regardless of the problem size. 
In fact, the maximum number of qubits,
\begin{equation}
    n_q = 2k + 1,
    \label{eqn:nq-l1}
\end{equation}
are required to simulate the expectation of a Hamiltonian with $k$-local observables, for this kind of ansatz (one layer of single-qubit gates and circular entanglement). 
The architecture of the entanglement is crucial in deciding how many qubits we can reduce.
For linear entanglement (only adjacent qubits entangled, first and last not entangled), there would be another case of the subcircuits where the observable is located
on the first qubit or last qubit. 
For full entanglement (all qubits entangled), LCC will not be possible. 


\begin{figure*}[t]
    \centering
    \includegraphics[width=\textwidth]{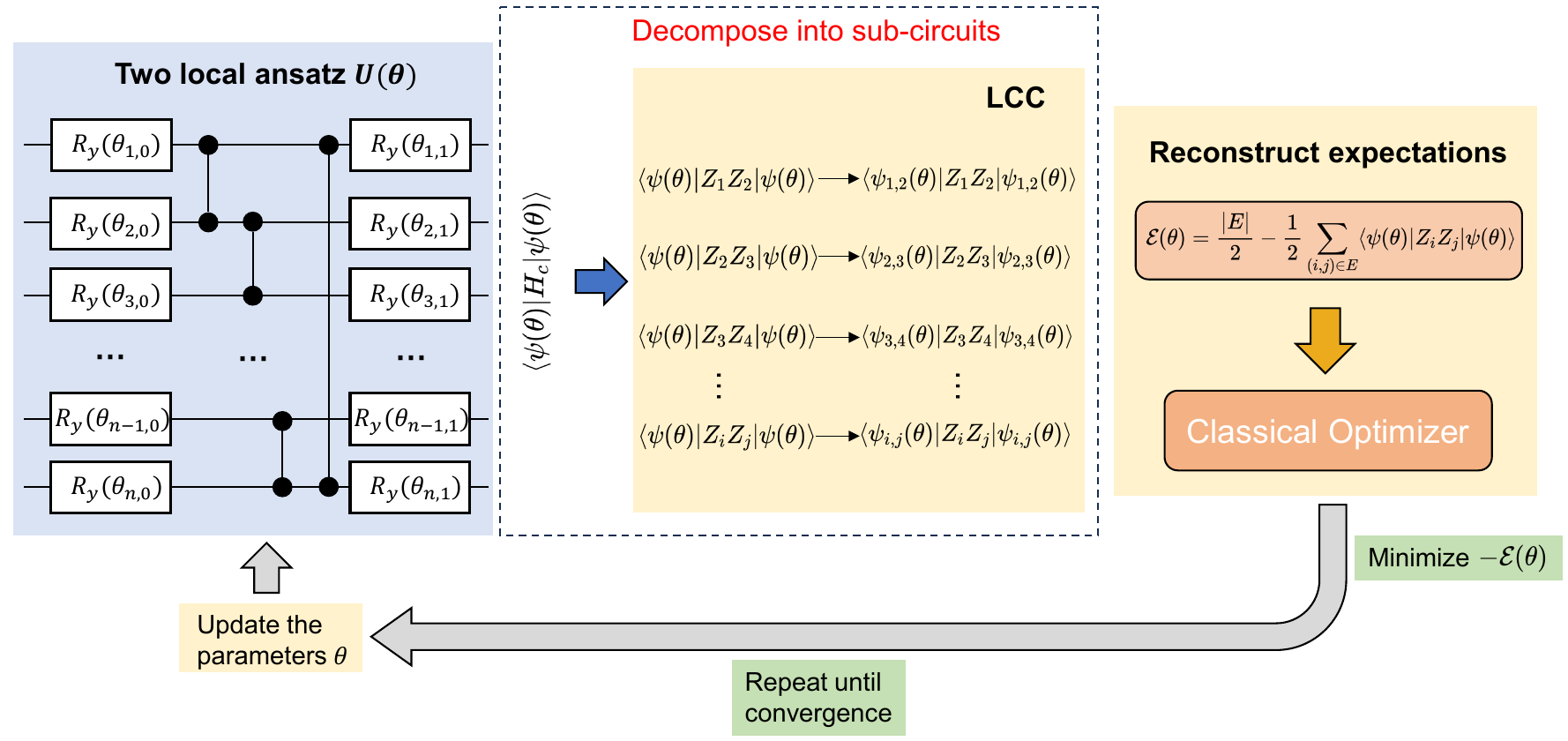}
    \caption{\justifying
    Overall workflow of the LCC in two-local ansatz. The expectation function $\expval{H_c}{\WF}$ of the original circuit is decomposed into separate expectation values of subcircuits using Light Cone Cancellation. These sub-expectations are used to reconstruct the total expectation function $\mathcal{E}(\bm{\theta})$ of the Max-Cut instances, which is then iteratively optimized within a hybrid quantum-classical framework.}
    \label{fig:flow-chart}  
\end{figure*}

Another advantage of LCC is that the number of gates is reduced, which in turn reduces the effect of gate noise in quantum devices. However, it is also worth noting that even though the numbers of qubits and gates are reduced, the number of parameters remains unchanged after LCC.
This is because the subcircuits after LCC will have different parameters corresponding to the indices $i$ and $j$,
depending on where the observables $Z_iZ_j$ are.
Therefore, the difficulty in parameter optimization remains the same before and after LCC.

LCC is also applicable for other circuits like QAOA and ma-QAOA. However, QAOA or ma-QAOA usually needs more layers (larger circuit depths) to achieve
higher ARs. 
Meanwhile, VQE with even one layer is enough to reach most of the states and hence easier to reach higher ARs than QAOA. 
This is due to the difference in expressibility between the VQE ansatz and the QAOA ansatz~\cite{expressibility}.
Since VQE yields higher ARs with fewer layers than QAOA, LCC-VQE can be done with fewer qubits, compared to LCC-QAOA or LCC-ma-QAOA.

Fig.~\ref{fig:flow-chart} shows the overall workflow of the LCC in two-local ansatz. The ansatz $U(\bm{\theta})$ prepares a parameterized quantum state $\ket{\WF}$,
which is then used to evaluate the expectation function $\expval{H_C}{\WF}$ for the Max-Cut Hamiltonian $H_C$.
By applying the Light Cone Cancellation (LCC), the original expectation value is decomposed into a set of smaller subcircuit expectations,
$\bra{\psi_{i, j}(\boldsymbol{\theta})} Z_i Z_j \ket{\psi_{i, j}(\boldsymbol{\theta})}$, each involving only a subset of qubits associated with local interactions in the Max-Cut instances.
These local expectations are subsequently combined to reconstruct the total expectation function $\mathcal{E}(\bm{\theta})$,
which is optimized using a classical optimizer. The parameters $\bm{\theta}$ are then iteratively updated in a hybrid quantum-classical loop until convergence.

\begin{figure}
    \centering
    \includegraphics[width=\linewidth]{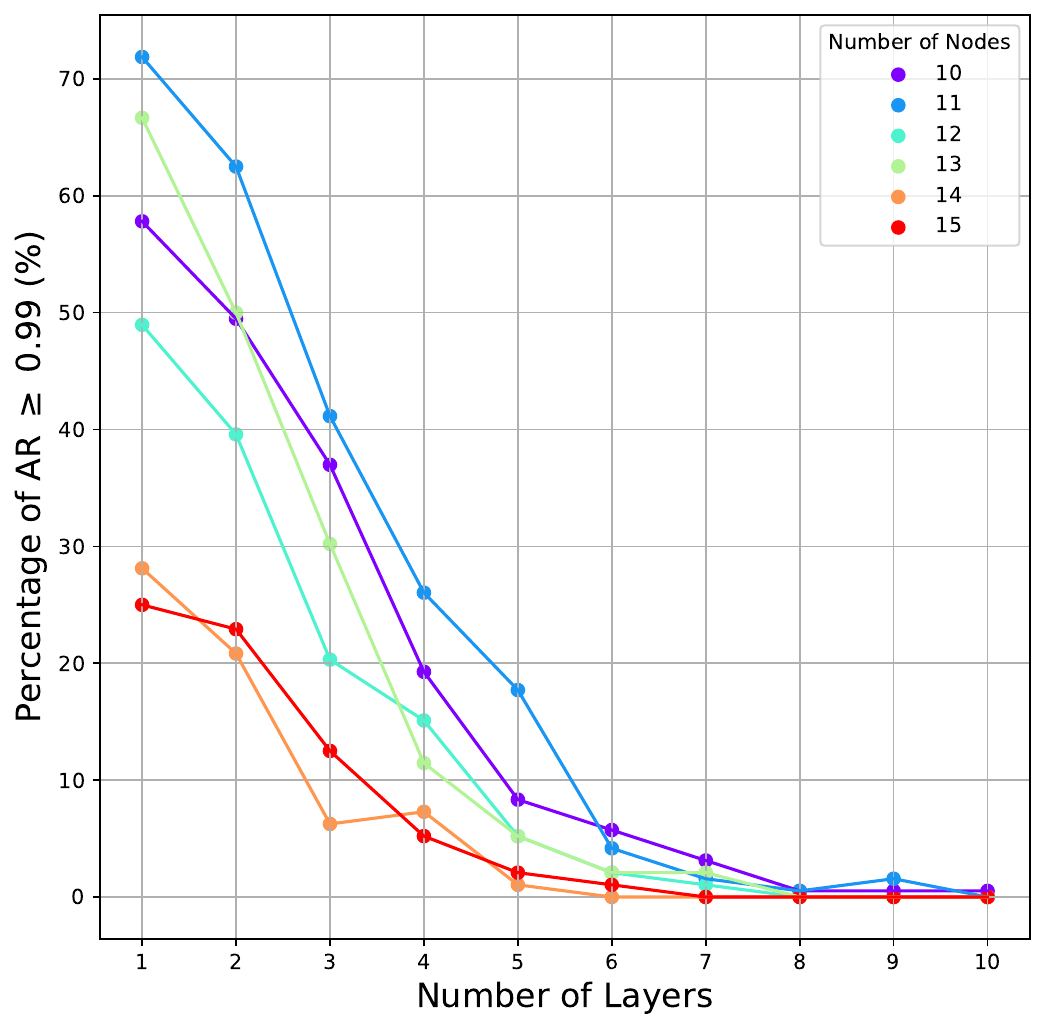}
    \caption{\justifying
    Percentage of AR $\geq 0.99$ in different number of layers of the ansatz.
     For $n = 10, 11, 12$, each point shows the percentage calculated from 8 graph instances, each with 24 different random initial parameters, with a total of 192 trials.
     For $n = 13, 14, 15$, each point shows the percentage for 4 graph instances, each with 24 trials, with a total of 96 trials.
     Each trial represents a set of random initial parameters, converged to the given AR after an optimization run.}
     
    \label{fig:ar-vs-layer}
\end{figure}

\subsection{Number of layers in the ansatz}
In this work, we only consider a single layer of the VQE ansatz as it is sufficient to address the Max-Cut problem.
To justify this, we conducted a simulation to investigate how the quality of the solution varies with the number of layers in the two-local ansatz.
Fig.~\ref{fig:ar-vs-layer} shows that as the number of layers increases, the chance of obtaining a high-quality solution (AR $\geq 0.99$) decreases. 
The chance of obtaining a high-quality solution is quantified by the percentage, i.e., the total number of times AR $\geq 0.99$ obtained, divided by the total number of trials. 
For $n=10,11,12$, each point shows the percentage calculated from 8 graph instances, each with 24 different random initial parameters, with a grand total of $8\times24 = 192$ trials. 
For $n=13,14,15$, each point shows the percentage for 4 graph instances, each with 24 trials, with a total of 96 trials.
The decline in the percentage can be explained by the increasing difficulty of optimization as the number of layers (number of parameters) increases, possibly the increased number of local minima, 
causing overparameterization where the AR could not increase further despite increasing the number of parameters.
Hence, we conclude that a single-layer ansatz is sufficient for the problem instances considered.

\begin{figure}
    \begin{flushleft}
    \resizebox{\linewidth}{!}{
    \begin{tikzpicture}
    \definecolor{lightblue}{RGB}{48, 102, 201} 
    \definecolor{darkgray}{RGB}{97, 105, 121} 
    
    \foreach \i in {0, 1, 2, 3} {
        \draw[gray] (\i, 0) -- (\i+1, 0); 
    }
    
    \foreach \i in {0, 1, 2, 3, 4} {
        \filldraw[lightblue] (\i, 0) circle (5pt); 
    }    
    \filldraw[darkgray] (2, 0) circle (5pt); 
    
    \node at (1, 0.4) {$Z_i$};
    \node at (3, 0.4) {$Z_j$};
    \node at (9, 0) {$n_q = 5$};
    \node at (-1, 0) {$L = 1$};
    
    \foreach \i in {0, 1, 2, 3, 4, 5, 6, 7} {
        \draw[gray] (\i, -2) -- (\i+1, -2); 
    }  
    
    \foreach \i in {0, 1, 2, 3, 4, 5, 6, 7, 8} {
        \filldraw[lightblue] (\i, -2) circle (5pt); 
    }
    
    \filldraw[darkgray] (4, -2) circle (5pt); 
    
    \node at (2, -1.6) {$Z_i$};
    \node at (6, -1.6) {$Z_j$};
    \node at (9, -2) {$n_q = 9$};
    \node at (-1, -2) {$L = 2$};
    
    \draw [decorate,decoration={brace,amplitude=5pt,mirror},yshift=-0.2cm] (0,-2.3) -- (1.9,-2.3) node [black,midway,yshift=-0.4cm] {$L$};
    \draw [decorate,decoration={brace,amplitude=5pt,mirror},yshift=-0.2cm] (2.1,-2.3) -- (3.9,-2.3) node [black,midway,yshift=-0.4cm] {$L$};
    \draw [decorate,decoration={brace,amplitude=5pt,mirror},yshift=-0.2cm] (4.1,-2.3) -- (5.9,-2.3) node [black,midway,yshift=-0.4cm] {$L$};
    \draw [decorate,decoration={brace,amplitude=5pt,mirror},yshift=-0.2cm] (6.1,-2.3) -- (8,-2.3) node [black,midway,yshift=-0.4cm] {$L$};
   
    \end{tikzpicture}
    }   
    \end{flushleft}
    \caption{\justifying
    The entanglement map after LCC when the number of ansatz layers $L$ increases. Each node represents a qubit and each edge represents an entanglement. 
    The figure shows the maximum number of qubits required for the linear or circular entanglement after LCC. The number of qubits required for LCC also depends on the number of qubits of the original ansatz,
    and also the distance between the observables. The maximum number of qubits is achieved when the observables are exactly $2L$ qubits apart of each other. 
    To achieve maximum number of qubits, each of the observables stretches out a distance of $L$ qubits (blue nodes) for two sides, and they overlap at the grey node, forming an inseparable entanglement.}
    \label{fig:layer-sg}
\end{figure}

Let us consider what happens to LCC when we have more than one layer of the ansatz. 
As the number of layers increases, the number of gates that can be cancelled decreases, resulting in a larger subcircuit after LCC.
This is because entangling gates (CZ gate in our case) farther from the center cannot commute through the layers nearer to the center (where the observables are) to get cancelled out
on the other side, causing them to remain in the circuit after LCC (refer to Fig.~\ref{fig:lcc}, where the area outside the light cone gets larger if the circuit has more layers).
The number of qubits remaining after LCC depends on the entanglement structure of the ansatz. 
The entanglement map of a quantum circuit can be viewed as a graph with the qubits as the nodes and the entanglement as the edges, e.g., if there is an entangling gate between qubit $i$ and qubit $j$, 
then there is an edge between node $i$ and node $j$.
The resulting entanglement map, after LCC, can then be viewed as a subgraph spanned by the observables $Z_i$ and $Z_j$, from the distance $L$ nodes away from node $i$ and node $j$, 
where $L$ is the number of layers in the original ansatz. 
This is analogous to the idea where QAOA is said to search deeper subgraphs as its circuit depth $p$ increases~\cite{farhi2014quantum,brandao2018fixed,galda2021transferability}.
Fig.~\ref{fig:layer-sg} shows the visualization of the entanglement map after LCC, considering the largest subgraph that can be spanned by the two observables.
For linear and circular entanglements, the subgraphs are represented by line graphs. 
To calculate the maximum number of qubits required after LCC, we consider the case where the size of the subgraph is at its maximum. 
This happens when the observable nodes stretch out for a distance to each side and overlap at their ends, similar to the subcircuit in Fig.~\ref{fig:lcc}(c) in the case of $L=1$. 
The grey nodes in Fig.~\ref{fig:layer-sg} show the overlapping nodes for each observable.
Thus, it is not difficult to establish a relation between the maximum number of qubits required $n_q$ with the number of layers $L$:
\begin{equation}
    n_q = 4L + 1,
    \label{eqn:nq-k2}
\end{equation}
for 2-local observables like the Max-Cut Hamiltonian. For $k$-local observables, the maximum number of qubits is 
\begin{equation}
    n_q = 2kL + 1.
    \label{eqn:nq-general}
\end{equation}
Note that $n_q$ is the maximal case where $n \geq n_q$ and the observables are exactly $2L$ qubits apart from each other.
The maximum number of qubits required will still be bounded by the original graph size $n$, and also when the observables are nearer or farther from each other. 
Eq.~(\ref{eqn:nq-l1}) and (\ref{eqn:nq-k2}) are the special cases for Eq.~(\ref{eqn:nq-general}) when $L=1$ and $k=2$ respectively, in the case where the ansatz only has one layer, or the observables are 2-local.

\begin{figure*}
    \centering
    \includegraphics[width=0.8\linewidth]{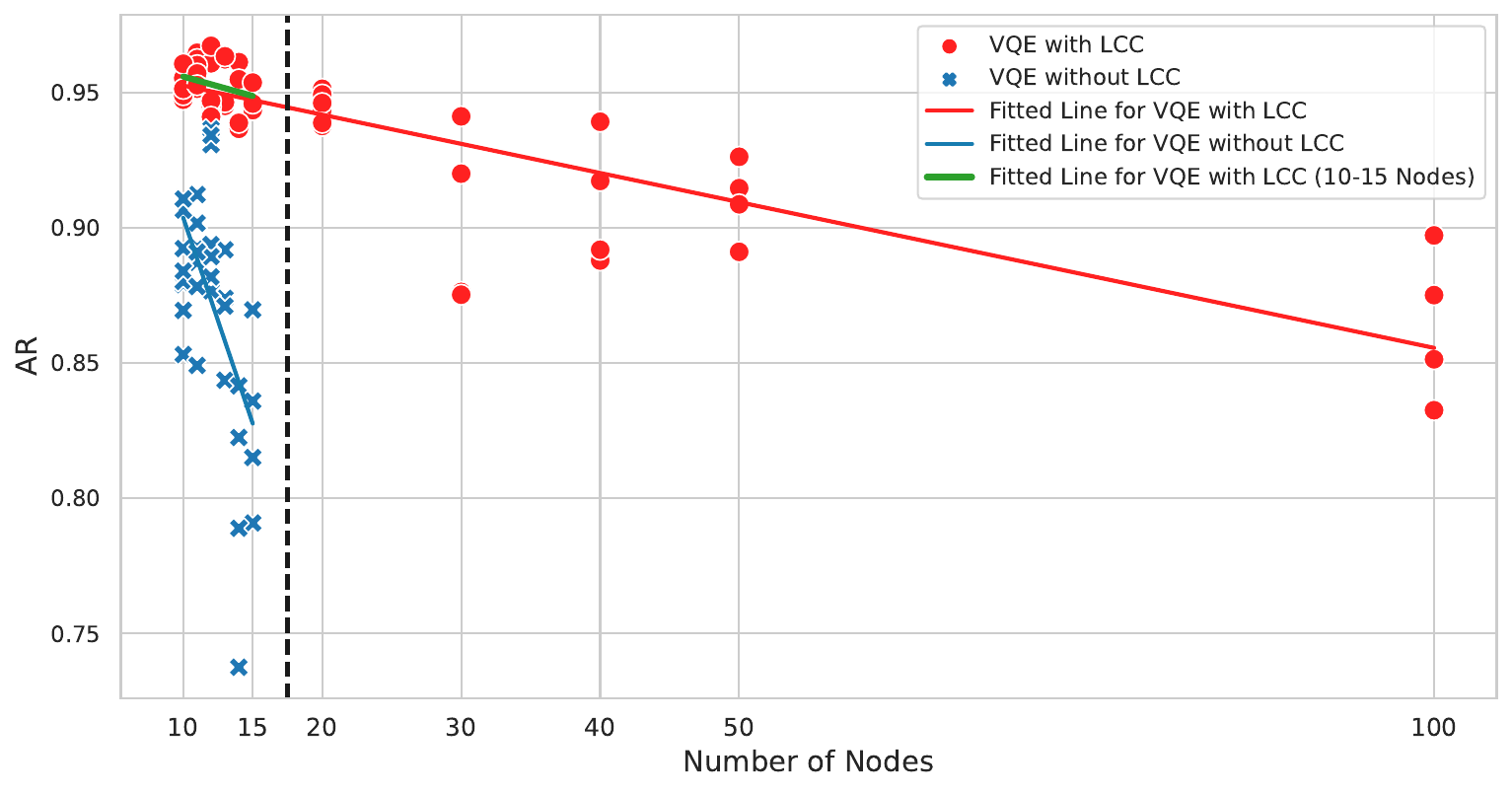}
    \caption{\justifying
    Comparison of the approximation ratio (AR) for the VQE solved with a 7-qubit fake backend \texttt{FakeCasablanca} (with LCC) and a %
    27-qubit fake backend \texttt{FakeParis} (without LCC). %
    Each point shows the best AR (highest) chosen out of 24 trials. The lines are linear fits of their respective data. Meanwhile, the gradients of the red line, blue line, and green line are $-0.0011$, $-0.0152$ and $-0.0018$, respectively.}
    \label{fig:7v27}
\end{figure*}

\section{Results}
\subsection{LCC under noisy conditions}
We solve the Max-Cut problem using VQE with a two-local ansatz (single layer $R_y$ and CZ entanglement). 
We employ the COBYLA optimizer~\cite{cobyla}, along with the AerSimulator provided by Qiskit~\cite{qiskit2024}, for all the simulations.
The demonstrations are designed to compare the performance of the VQE with LCC and that of VQE without LCC. 
We measure the performance using the approximation ratio (AR), which indicates how close the result given by VQE is to the optimal solution.
To make sure that the optimizer does not converge to a good minimum by chance, we perform 24 trials with random initial parameters for every instance, i.e., 
random initialization.
The demonstrations are performed under noisy conditions so that we can observe the effect of the reduction in the number of qubits and the number of gates 
on the amount of noise in the circuit. 
We use two different fake noisy devices provided by Qiskit~\cite{qiskit}: \verb+FakeCasablanca+ (7 qubits) and \verb+FakeParis+ (27 qubits). 
These fake devices simulate the same noise settings in their respective real quantum devices.
The specifications of the two noisy devices are stated in the Appendix~\ref{sec:fakebackend}.
We did two different comparisons to show the noise mitigation of LCC resulted by two different factors:
1) we compare the results for the 7-qubit vs. the 27-qubit devices to show that LCC allows us
to run the subcircuits on a smaller device, which results in noise mitigation;
2) we then compare the execution of LCC on the same 27-qubit device to show the noise mitigation due to the reduced number of gates.

\begin{figure}
    \centering
    \includegraphics[width=\linewidth]{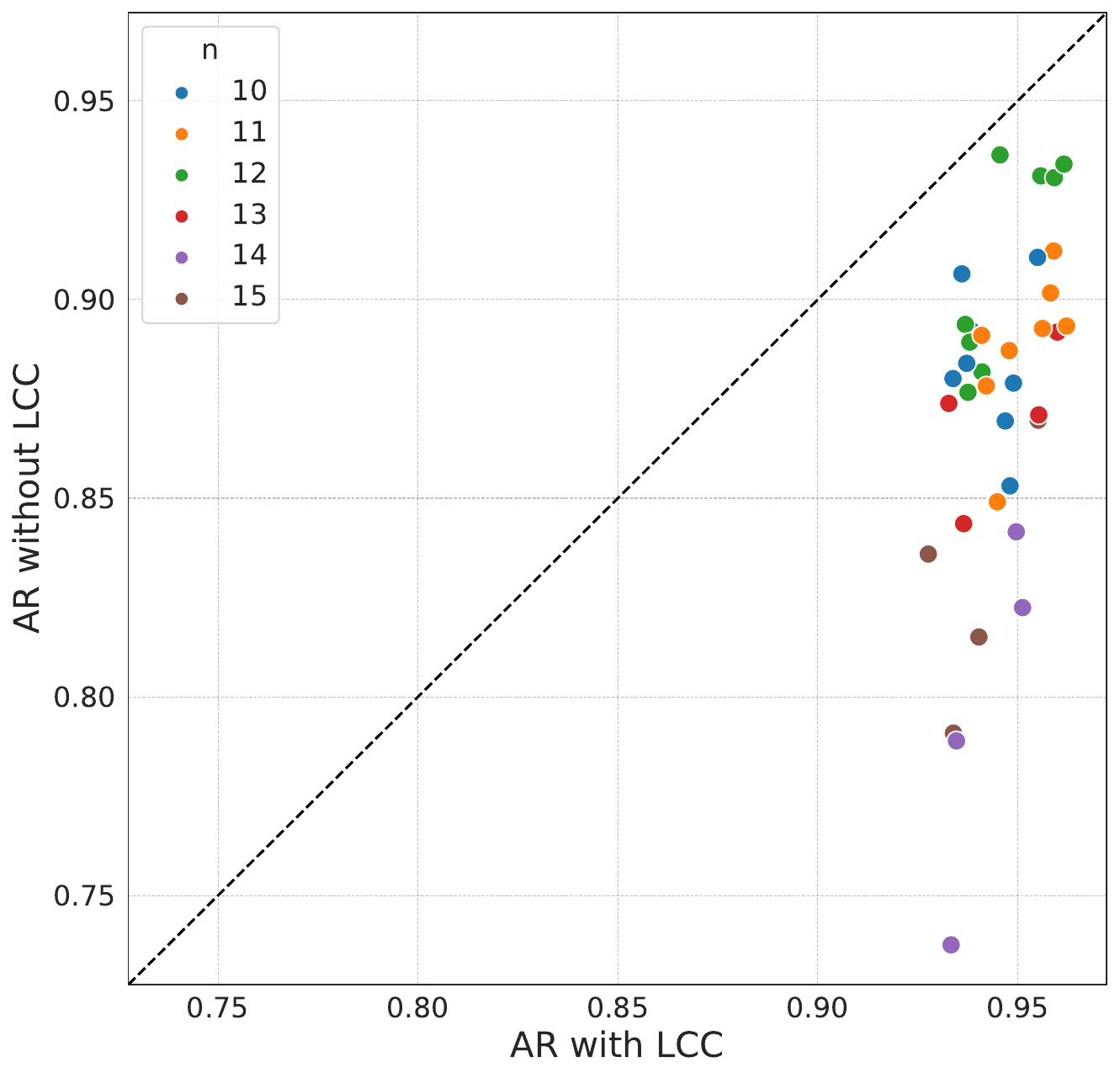}
    \caption{\justifying
    Comparison of the AR for the VQE with LCC and without LCC, using the same \texttt{FakeParis} backend (27 qubits). %
    The diagonal dashed line shows where the AR of both methods are equal. All the points are in the lower triangle and represent higher AR with LCC.}%
    \label{fig:27v27}
\end{figure}

Figure~\ref{fig:7v27} shows the comparison between Max-Cut instances solved with LCC on the 7-qubit fake backend, and those solved without LCC on the 27-qubit fake backend.
As only 5 qubits are required to simulate the subcircuits for LCC, the VQE simulation can be run on a device with 7 qubits. 
On the other hand, simulation with full number of qubits is required for those without LCC. 
The main purpose of this figure is to show the possibility of running LCC in a smaller device with less noise,
while Figure~\ref{fig:27v27} shows the comparison of the effect of noise on the same device, attributed to the reduced number of gates after LCC.
In both settings, we solve the Max-Cut for 36 non-isomorphic instances, ranging from number of vertices $n=10$ to $n=15$. 
Additonally, 24 non-isomorphic instances are solved on the 7-qubit fake backend (with LCC) for $n=20,30,40,50$ and $100$.
The dataset for the demonstrations is shown in the Appendix~\ref{sec:dataset}.
Each point in the plot represents the best AR out of 24 trials for a single problem instance. 
The red points plot the ARs for the instances solved with LCC on the 7-qubit fake backend; 
the blue points plot the ARs for the instances solved without LCC on the 27-qubit fake backend. 
The red line is a linear fit through the red points (with LCC) for $n=10$ to $n=100$.
The blue line linearly fits through the blue points (without LCC) for $n=10$ to $n=15$. 
The green line is a fit for the red points (with LCC) from $n=10$ to $n=15$.

There are a few observations worth noting. LCC enabled the simulation of large problems up until $n=100$, only with quantum circuits with at most 5 qubits. 
From the red and blue fitted lines, we can see that the ARs for problems with LCC are generally higher than those without LCC. 
Although with different environments (7-qubit and 27-qubit fake backends), the error rate is generally lower on a smaller device, so the AR is not so much 
deteriorated on a 7-qubit device. Moreover, with less number of gates, the effect of noise on the AR is also reduced. 
It can also be observed that the AR decreases as the problem size (number of the graph vertices) increases. 
Another interesting point to note is that the green line has similar slope as the red line, which means the decreasing trend is similar for 
microscopic ($n=10$ to $n=15$) and macroscopic ($n=10$ to $n=100$) number of nodes. 
Also, we can observe from the blue line that if without LCC, the AR decreases faster due to more noises in the circuits. 
This implies the possibility that in the case without LCC, the AR would decrease faster than the case with LCC, if the blue line is extrapolated
to larger problem size.

\begin{figure*}[t]
    \centering
    \begin{subfigure}[b]{0.48\linewidth}
        \centering
        \caption*{(a)}
        \vspace{1ex}
        \includegraphics[width=\linewidth]{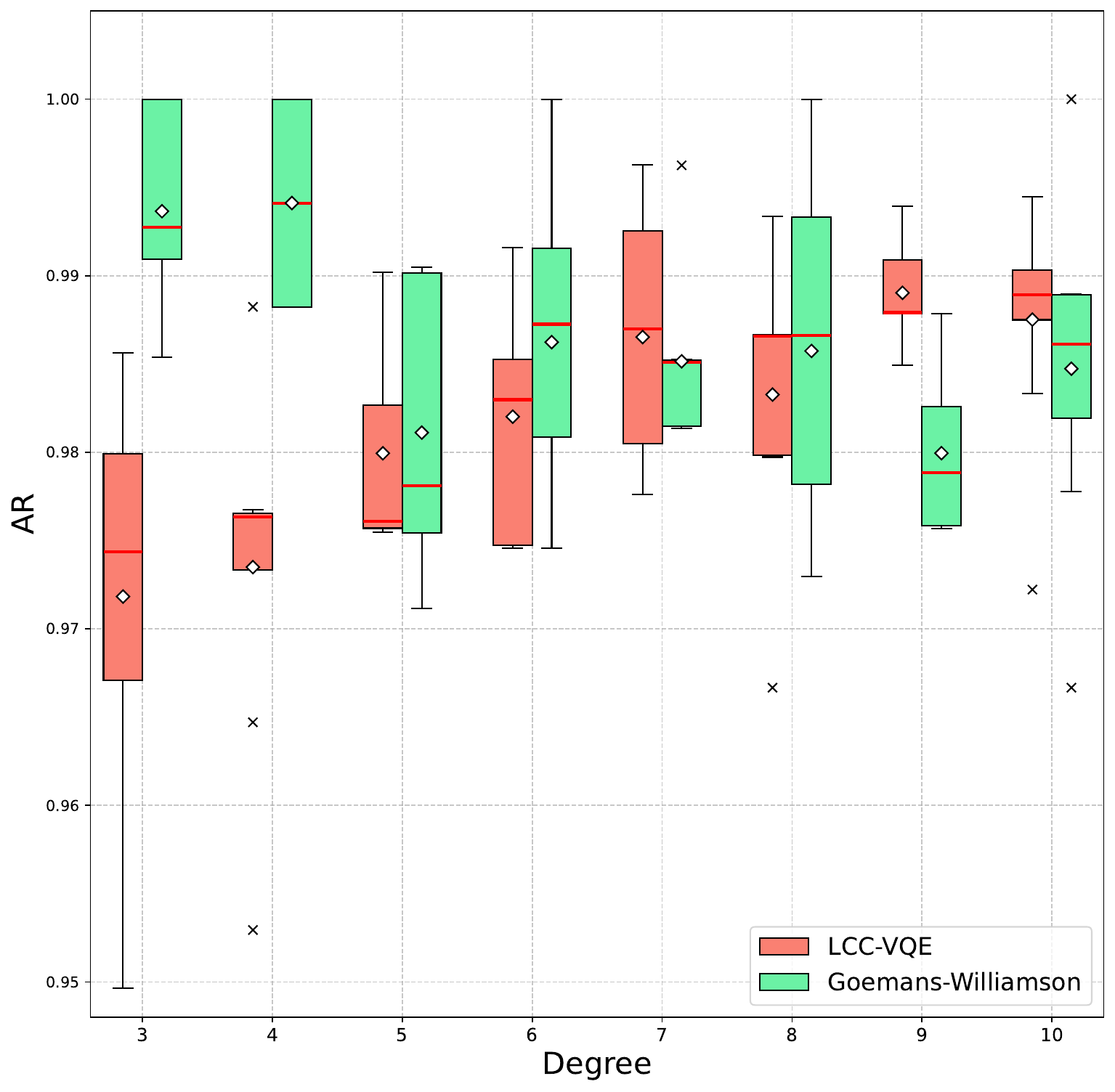}
        \label{fig:GW_LCC100reg}
    \end{subfigure}\hfill
    \begin{subfigure}[b]{0.48\linewidth}
        \centering
        \caption*{(b)}
        \vspace{1ex}
        \includegraphics[width=\linewidth]{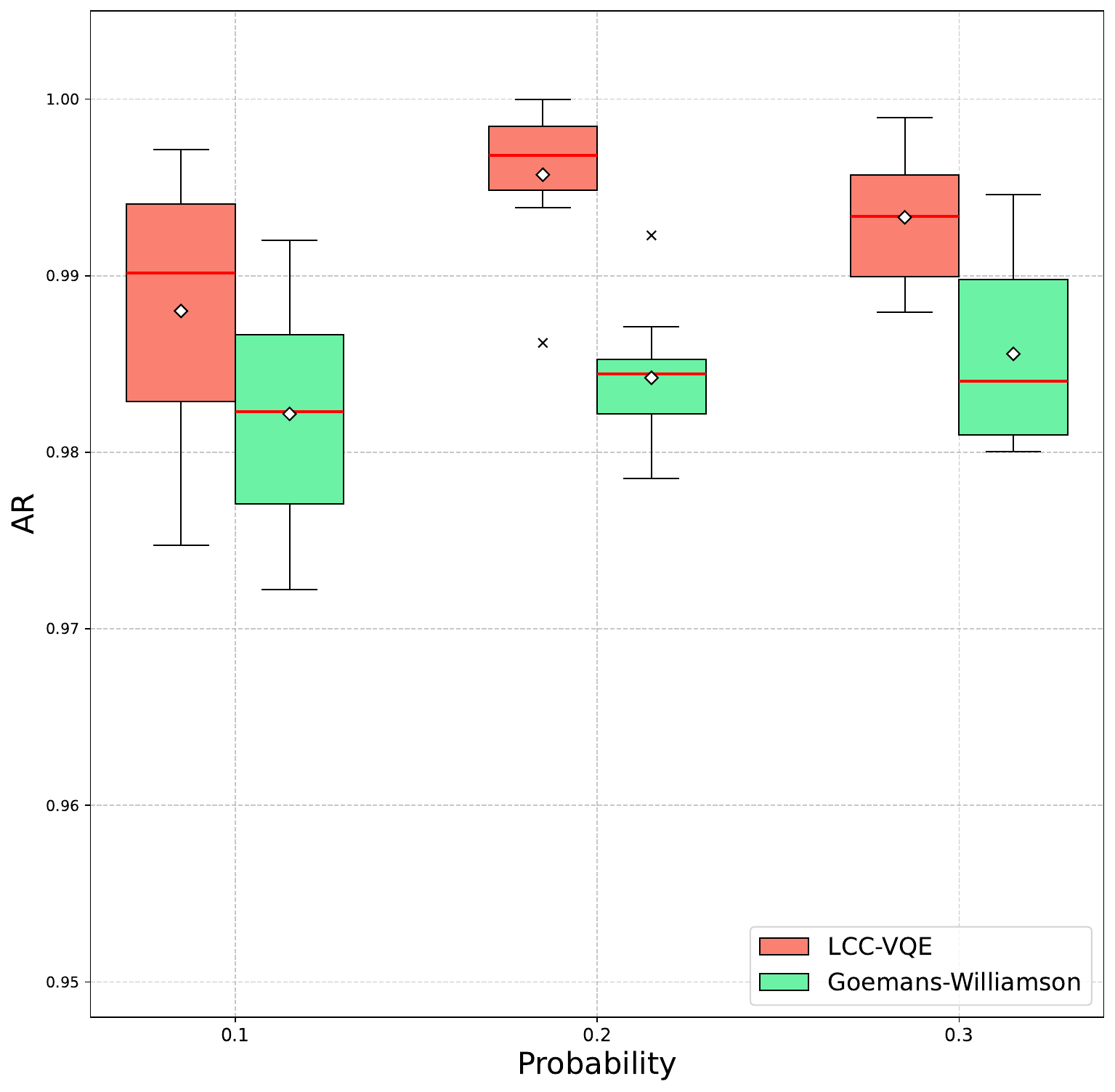}
        \label{fig:GW_LCC100gnp}
    \end{subfigure}
    \caption{\justifying (a) Comparison of the AR for LCC-VQE and GW algorithm on 100-vertex $d$-regular graph instances, relative to the exact solution found by GUROBI. Each degree corresponds to a set of 8 different instances. The boxplots of both algorithms show the best AR chosen out of 24 trials for each instance, with 8 instances at each degree.
    (b) Comparison of the AR for LCC-VQE and GW algorithm on 100-vertex $G(n,p)$ graph instances. For each edge probability $p = 0.1, 0.2, 0.3$, we generate 8 graph instances. The boxplots show the best AR for each instance, selected from 24 trials for LCC-VQE and the GW algorithm.
    The red lines and white diamonds of boxplots represent the medians and means respectively.}
    \label{fig:GW_LCC}
\end{figure*}

Figure~\ref{fig:27v27} shows the comparison for $n=10$ to $n=15$ problems solved with and without LCC, using the same 27-qubit backend. 
The 36 instances used are the same as those in Fig.~\ref{fig:7v27}.
This figure is to show the effect of noise mitigation after LCC attributed to the reduction in the number of gates.
The diagonal dashed line shows where both methods (with and without LCC) have the same AR.
All the points are in the lower triangle, which represents that LCC gives higher ARs than without having LCC. 
Under the same noise conditions, it can be observed that all 36 instances give higher ARs with LCC applied.
This result shows the effect of the reduction in the number of gates in a more evident way than the demonstration shown in Fig.~\ref{fig:7v27},
as the number of qubits and the error rates are the same for both with and without LCC.
It is also observed that problems with larger sizes $n$ benefit more from LCC as their ARs stay away from the diagonal dashed line, 
and those with smaller sizes stay near the diagonal dashed line.
This is because larger circuits generally have more noise, causing their ARs to deteriorate more.

\subsection{LCC vs. Goemans-Williamson (GW) algorithm}
It is worth noting that under noisy conditions, although the solutions found by LCC-VQE are capable of approximating the true solutions of problems, they consistently fall short of those achieved by the GW algorithm.
The GW algorithm serves as a typical benchmark for classical approximation algorithms for the Max-Cut problem.
To better evaluate the inherent potential of LCC-VQE without the influence of noise, we therefore conduct further comparative simulations under noiseless conditions.
Fundamentally, according to Eq.~(\ref{eqn:lcc-equals-no-lcc}), the comparison is essentially between the VQE algorithm and the GW algorithm, regardless of whether LCC is applied.
Previous studies~\cite{munoz2024low} have observed that the performance of the GW algorithm deteriorates as the problem size increases.
Furthermore, numerical simulations of XQAOA and the GW algorithm on 128- and 256-vertex regular graphs have shown that the AR of GW tends to decrease as the degree of the graph increases~\cite{vijendran2024expressive}.

Motivated by these observations, we seek to establish an effective benchmarking strategy for evaluating the performance of LCC-VQE by comparing it against GW on large size problem instances.
Consequently, we construct another dataset where all instances have $100$ vertices but different degrees. The subsequent comparative simulations employ LCC in a noiseless environment using the StatevectorEstimator provided by Qiskit, and the circuit remains previous two-local ansatz.
To obtain the exact Max-Cut values and compute AR in these simulations, we employ the GUROBI solver~\cite{gurobi}, which is widely used in industry.
Fig.~\ref{fig:GW_LCC}(a) shows the comparison of AR for both algorithms on 100-vertex $d$-regular graph instances with degrees $d$ ranging from 3 to 10, giving a total of 64 instances. Each boxplot represents the best ARs across 8 instances at each degree, where each best AR is selected from 24 trials of each instance. 
The result shows that the GW algorithm demonstrates consistently strong performance on low-degree regular instances ($d =3, 4$), with the majority of ARs in the boxplots exceed 0.99 and are close to 1.
However, although the overall performance of GW declines as the degree increases, the trend is neither gradual nor monotonic. In particular, the lowest median AR across all degrees is observed at $d = 5$.
In contrast, LCC-VQE exhibits a steadily increasing trend in ARs as the degree increases, and its median values gradually improve as the degree increases. It begins to outperform GW in the upper quartile, median, and lower quartile values on graphs with degree $d \geq 9$.

Then, we conduct the comparative simulation for LCC-VQE and the GW algorithm on 100-vertex \ER (also known as $G(n,p)$) graph instances.
Consistent with our previous work, we employ the COBYLA optimizer to optimize the parameters in this simulation.
Fig.~\ref{fig:GW_LCC}(b) presents the comparison of the AR across graph instances with different edge probabilities $p$.
For each probability $p = 0.1, 0.2, 0.3$, 8 graph instances are generated, giving a total of 24 instances. 
Each boxplot represents the best ARs across 8 instances at each edge probability, where for each instance the best AR is selected from 24 trials for both LCC-VQE and the GW algorithm.

As shown in the Fig.~\ref{fig:GW_LCC}(b), LCC-VQE surpasses GW in terms of the upper quartile, lower quartile, mean, and median ARs at different edge probability $p$. Notably, the number of edges in the instances at $p = 0.1$ is comparable to that in Fig.~\ref{fig:GW_LCC}(a) for graphs with degree $d = 10$.
The number of edges in regular graphs is $nd/2$, whereas the average number of edges in $G(n,p)$ graphs is $pn(n-1)/2$.
Despite this similarity, the upper quartile and median ARs of LCC-VQE at $p = 0.1$ are higher than that at $d = 10$. As the edge probability increases to $p = 0.2$, the AR of LCC-VQE further increases. Although a slight drop is observed at $p = 0.3$, the majority of the box remains above 0.99.

Fig.~\ref{fig:GW_LCC}(a) and (b) show similar trend in the comparison of XQAOA and GW in~\cite{vijendran2024expressive},
in which LCC-VQE (or XQAOA) initially shows lower AR than GW at small degrees, but surpasses the performance of GW as the degree increases.

\section{Conclusion}
In this work, we presented the LCC on VQE and studied what the effect it acts in solving the Max-Cut problem.
Our work opens up the possibility of using VQE to solve combinatorial optimization problems, 
as VQE requires less number of layers than QAOA to achieve the same performance. This allows us to cancel a larger number of qubits when applying LCC,
thereby shifting the complexity of circuit simulation from exponential scaling (number of qubits) to polynomial scaling (number of edges in a graph). 
For the Max-Cut problem with a two-local cost Hamiltonian, only at most five qubits are required to solve the problem of any size.
Concerning the implementation of LCC, our preliminary calculations reveal the relationship between the maximum number of qubits $n_q$ with $k$-local observables in calculating the expectation of Hamiltonian $H_C$. It is worth noting that LCC can only be implemented under linear and circular entanglement structures, whereas it is not feasible under full entanglement.
To look at the precise relation between high-quality solution and layer numbers in addressing Max-Cut problem, 
we compare the performance of VQE ansatz with different number of layers and conclude that the opportunity to achieve a high-quality solution (AR $\geq 0.99$) comes to decrease with the number of layers $L$ increases. This decline is attributed to tendency to become trapped in local minimas due to overparameterization. Meanwhile, the computational cost increases with the number of layers $L$, and those causes make it necessary for the optimization process to achieve faster convergence rate by setting the ansatz to a single layer.
Furthermore, as the number of layers $L$ increases, the size of subcircuits with the maximum number of qubits $n_q = 4L +1 $ also grows. 
This can significantly undermine the effectiveness of LCC in a noisy environment.

We compare the performance of circuits with and without LCC on a noisy simulator provided by Qiskit. 
The results show that the circuits with LCC generally yield higher approximation ratios than the circuits without LCC, hence implying that the noise is being mitigated.
Furthermore, the performance of LCC-VQE and the GW algorithm is compared under a noiseless condition. The results show that even on denser graphs, LCC-VQE exhibits significant advantages and potential.
As presented in the previous sections, it is unclear whether the LCC could lead to a quantum advantage for VQE in a lower layer with circular entanglement. To overcome those challenges, one direction is to conduct more comprehensive and in-depth simulations. 
It would be fruitful to pursue further research about combining LCC with other algorithms in order to reduce circuit's noise.
Additionally, due to its advantages with fewer layers, LCC-VQE is hoped to be tested on other combinatorial optimization problems.
It could become an interesting and valuable development for helping VQA mitigate the effects of noise.

\section*{Acknowledgments}
This work was supported by JST SPRING, Grant Number JPMJSP2124 and by the National Research Foundation, Singapore under its Quantum Engineering Programme 2.0 (NRF2021-QEP2-02-P01).

\section*{Data availability}
We provide the code that implements the LCC framework, along with the benchmark and simulation datasets used in this paper.
These resources will be made publicly available at \href{https://github.com/xenoicwyce/lcc}{https://github.com/xenoicwyce/lcc}~\cite{lcc_code} after this work is published.

\section*{Author contributions}
All authors contributed to the work enclosed in the paper and the writing of the paper.

\appendix

\section{Dataset of the simulations}\label{sec:dataset}
Table~\ref{tab:dataset_noise} and Table~\ref{tab:dataset_noiseless} show the simulation datasets used in this study. We consider two categories of unweighted, undirected graphs: the $G(n,p)$ \ER graphs and the regular graphs.
The graphs are generated using the \verb+NetworkX+ Python package. The $G(n,p)$ graphs are generated using \verb+fast_gnp_random_graph()+; 
the regular graphs are generated using \verb+random_regular_graph()+, with the seeds specified.
Each seed represents one graph instance.

\renewcommand{\arraystretch}{1.5}
\begin{table}[ht]
\centering
\caption{\justifying
The datasets used in the comparative demonstrations of two fake backends---the regular graphs with degree $d$ and $G(n,p)$ graphs with edge probability $p$.}
\resizebox{0.5\textwidth}{!}{%
\begin{tabular}{|c|c|c|c|}
\hline
\textbf{No. of nodes} & \textbf{Graph type} & \textbf{$d$ (Reg.) or $p$ (ER)} & \textbf{Seed}\\ \hline
10 & ER & 0.5 & 0, 1, 2, 3 \\ \hline
10 & Reg. & 3 & 0, 1, 2, 3 \\ \hline
11 & ER & 0.5 & 0, 1, 2, 3 \\ \hline
11 & Reg. & 4 & 0, 1, 2, 3 \\ \hline
12 & ER & 0.5 & 0, 1, 2, 3 \\ \hline
12 & Reg. & 3 & 0, 1, 2, 3 \\ \hline
13 & ER & 0.5 & 0, 1 \\ \hline
13 & Reg. & 2 & 1, 3 \\ \hline
14 & ER & 0.4 & 0, 1 \\ \hline
14 & Reg. & 2 & 2, 5 \\ \hline
15 & ER & 0.3 & 1, 2 \\ \hline
15 & Reg. & 2 & 3, 7 \\ \hline
20 & ER & 0.25 & 0, 3, 5, 17 \\ \hline
20 & Reg. & 3 & 0, 1, 2, 3 \\ \hline
30 & ER & 0.12 & 0, 5 \\ \hline
30 & Reg. & 2 & 3, 8 \\ \hline
40 & ER & 0.06 & 106, 125 \\ \hline
40 & Reg. & 2 & 0, 1 \\ \hline
50 & ER & 0.06 & 126, 167, 424, 561 \\ \hline
100 & ER & 0.035 & 86520, 769454 \\ \hline
100 & Reg. & 3 & 0, 1 \\ \hline
\end{tabular}
}
\label{tab:dataset_noise}
\end{table}

\renewcommand{\arraystretch}{1.5}
\begin{table}[ht]
    \centering
    \caption{\centering 
    The datasets used in the comparative simulations of LCC-VQE and GW algorithm under a noiseless condition.}
    \resizebox{0.48\textwidth}{!}{%
    \begin{tabular}{|c|c|c|c|}
    \hline
    \textbf{No. of nodes} & \textbf{Graph type} & \textbf{$d$ (Reg.) or $p$ (ER)} & \textbf{Seed}\\ \hline
    100 & Reg. & 3 & \ 0, 1, 2, \ldots, 7\ \\ \hline
    100 & Reg. & 4 & \ 0, 1, 2, \ldots, 7\ \\ \hline
    100 & Reg. & 5 & \ 0, 1, 2, \ldots, 7\ \\ \hline
    100 & Reg. & 6 & \ 0, 1, 2, \ldots, 7\ \\ \hline
    100 & Reg. & 7 & \ 0, 1, 2, \ldots, 7\ \\ \hline
    100 & Reg. & 8 & \ 0, 1, 2, \ldots, 7\ \\ \hline
    100 & Reg. & 9 & \ 0, 1, 2, \ldots, 7\ \\ \hline
    100 & Reg. & 10 & \ 0, 1, 2, \ldots, 7\ \\ \hline
    100 & ER & 0.1 & \ 0, 1, 2, \ldots, 7\ \\ \hline
    100 & ER & 0.2 & \ 0, 1, 2, \ldots, 7\ \\ \hline
    100 & ER & 0.3 & \ 0, 1, 2, \ldots, 7\ \\ \hline
    
    \end{tabular}}
    \label{tab:dataset_noiseless}
\end{table}

\section{Device specifications for the fake backends}\label{sec:fakebackend}
\begin{figure}
    \centering
    \includegraphics[width=0.5\linewidth]{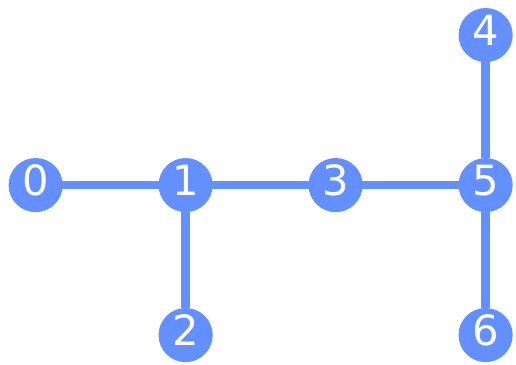}
    \caption{\centering Coupling map of \texttt{FakeCasablanca} (7 qubits).}
    \label{fig:fake_casablanca}
\end{figure}

\begin{figure}
    \centering
    \includegraphics[width=\linewidth]{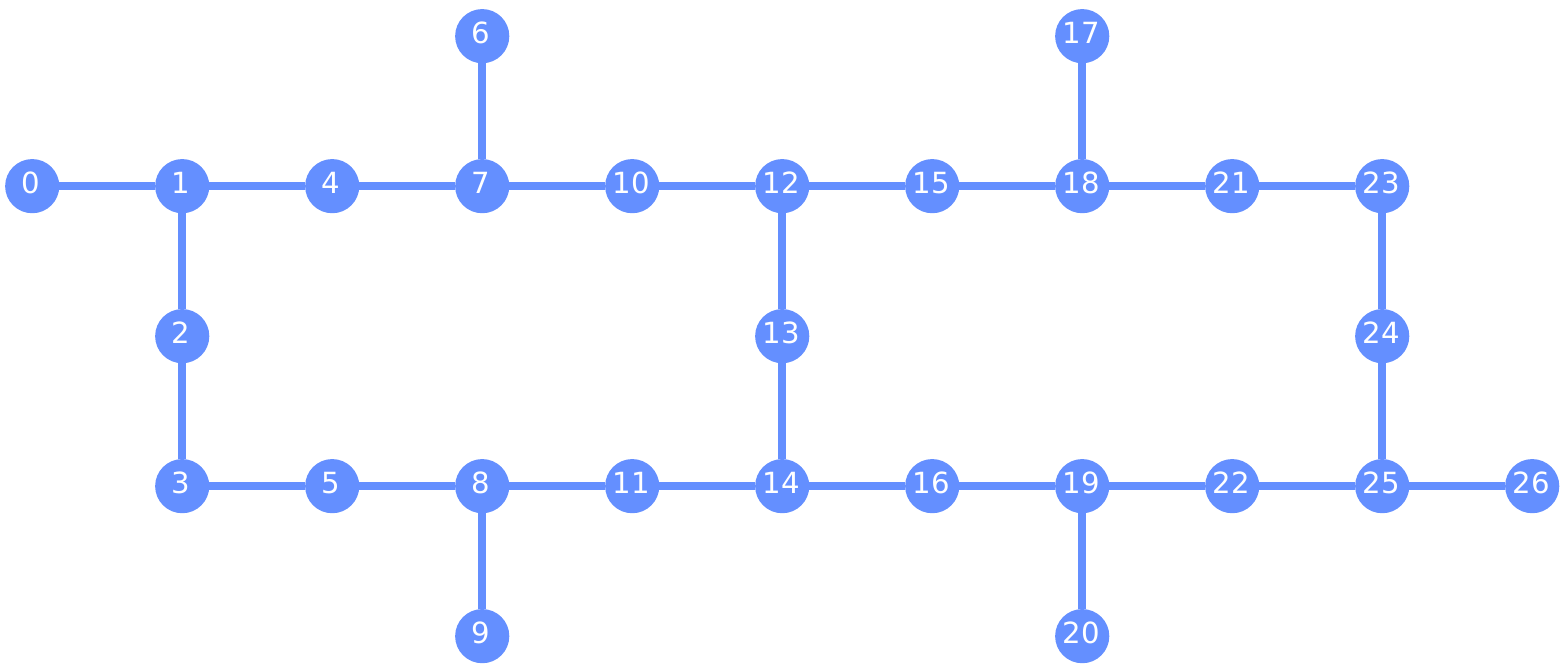}
    \caption{\centering Coupling map of \texttt{FakeParis} (27 qubits).}
    \label{fig:fake_paris}
\end{figure}

This section shows the detailed noise models for the two fake backends used in our demonstrations:
\texttt{FakeCasablanca} (7 qubits) and \texttt{FakeParis} (27 qubits).
The basis gates for the two fake devices are the $X$, $\sqrt{X}$, $R_z$ (rotational $Z$ gate) and the CNOT gate.
The $R_z$ gates have zero error. 
Fig.~\ref{fig:fake_casablanca} and \ref{fig:fake_paris} show the respective coupling maps for \texttt{FakeCascablanca} and \texttt{FakeParis}.
The CNOT errors for each coupling are shown in Table~\ref{tab:cnot_error_7qubits} and \ref{tab:cnot_error_27qubits}.
Table~\ref{tab:qubit_fake_casablanca} and \ref{tab:qubit_fake_paris} show the qubit characteristics for the fake backends
\texttt{FakeCascablanca} and \texttt{FakeParis} respectively.
The characteristics listed are the qubit frequencies, $T_1$ (relaxation time), $T_2$ (dephasing time), the readout errors,
and the single-qubit gate errors (for $X$ and $\sqrt{X}$ gates).

\clearpage 

\begin{table}
    \centering
    \caption{\centering CNOT errors on each qubit coupling for \texttt{FakeCasablanca} (7 qubits).}
    \label{tab:cnot_error_7qubits}
    \begin{tabular}{|c|c|}
      \hline
      \textbf{Qubit coupling} & \textbf{CNOT error} \\
      \hline
      (0,1) & $1.83\times10^{-2}$ \\ \hline
      (1,2) & $1.62\times10^{-2}$ \\ \hline
      (1,3) & $7.49\times10^{-3}$ \\ \hline
      (3,5) & $1.26\times10^{-2}$ \\ \hline
      (4,5) & $1.31\times10^{-2}$ \\ \hline
      (5,6) & $8.70\times10^{-3}$ \\ \hline
    \end{tabular}
\end{table}

\begin{table}
    \centering
    \caption{\centering CNOT errors on each qubit coupling for \texttt{FakeParis} (27 qubits).}
    \label{tab:cnot_error_27qubits}
    \begin{tabular}{|c|c|}
    \hline
    \textbf{Qubit coupling} & \textbf{CNOT error} \\ \hline
    (0,1)   & $1.26\times10^{-2}$ \\ \hline
    (1,2)   & $1.42\times10^{-2}$ \\ \hline
    (1,4)   & $1.24\times10^{-2}$ \\ \hline
    (2,3)   & $6.52\times10^{-2}$ \\ \hline
    (3,5)   & $1.40\times10^{-2}$ \\ \hline
    (4,7)   & $1.36\times10^{-2}$ \\ \hline
    (5,8)   & $1.29\times10^{-2}$ \\ \hline
    (6,7)   & $1.11\times10^{-2}$ \\ \hline
    (7,10)  & $1.03\times10^{-2}$ \\ \hline
    (8,9)   & $9.56\times10^{-3}$ \\ \hline
    (8,11)  & $7.75\times10^{-3}$ \\ \hline
    (10,12) & $1.86\times10^{-2}$ \\ \hline
    (11,14) & $1.20\times10^{-2}$ \\ \hline
    (12,13) & $1.71\times10^{-2}$ \\ \hline
    (12,15) & $1.29\times10^{-2}$ \\ \hline
    (13,14) & $1.18\times10^{-2}$ \\ \hline
    (14,16) & $1.96\times10^{-2}$ \\ \hline
    (15,18) & $2.40\times10^{-2}$ \\ \hline
    (16,19) & $1.32\times10^{-2}$ \\ \hline
    (17,18) & $2.92\times10^{-2}$ \\ \hline
    (18,21) & $1.42\times10^{-2}$ \\ \hline
    (19,20) & $1.32\times10^{-2}$ \\ \hline
    (19,22) & $1.27\times10^{-2}$ \\ \hline
    (21,23) & $1.39\times10^{-2}$ \\ \hline
    (22,25) & $7.60\times10^{-3}$ \\ \hline
    (23,24) & $1.75\times10^{-2}$ \\ \hline
    (24,25) & $8.79\times10^{-3}$ \\ \hline
    (25,26) & $8.01\times10^{-3}$ \\ \hline
    \end{tabular}
\end{table}

\begin{table*}
    \centering
    \caption{\centering Qubit characteristics for \texttt{FakeCasablanca} (7 qubits).}
    \label{tab:qubit_fake_casablanca}
    \begin{tabular}{|c|c|c|c|c|c|}
      \hline
      \textbf{Qubit} & \textbf{Frequency {\scriptsize (GHz)}} & \textbf{$T_1\,(\mu s)$} & \textbf{$T_2\,(\mu s)$} & \textbf{Readout error} & \textbf{$X$ and $\sqrt{X}$ gate error} \\
      \hline
      0 & $4.82$ & $102.88$ & $53.56$  & $1.97\times10^{-2}$ & $3.51\times10^{-4}$ \\ \hline
      1 & $4.76$ & $111.41$ & $109.47$ & $1.44\times10^{-2}$ & $8.98\times10^{-4}$ \\ \hline
      2 & $4.91$ & $78.93$  & $132.42$ & $1.78\times10^{-2}$ & $3.62\times10^{-4}$ \\ \hline
      3 & $4.88$ & $88.06$  & $77.27$  & $1.56\times10^{-2}$ & $2.61\times10^{-4}$ \\ \hline
      4 & $4.87$ & $98.93$  & $39.94$  & $1.54\times10^{-2}$ & $5.43\times10^{-4}$ \\ \hline
      5 & $4.96$ & $83.95$  & $122.30$ & $1.74\times10^{-2}$ & $4.81\times10^{-4}$ \\ \hline
      6 & $5.18$ & $65.59$  & $63.50$  & $3.26\times10^{-2}$ & $4.63\times10^{-4}$ \\ \hline
    \end{tabular}
\end{table*}

\begin{table*}
    \centering
    \caption{\centering Qubit characteristics for \texttt{FakeParis} (27 qubits).}
    \label{tab:qubit_fake_paris}
    \begin{tabular}{|c|c|c|c|c|c|}
      \hline
       \textbf{Qubit} & \textbf{Frequency {\scriptsize (GHz)}} & \textbf{$T_1\,(\mu s)$} & \textbf{$T_2\,(\mu s)$} & \textbf{Readout error} & \textbf{$X$ and $\sqrt{X}$ gate error} \\
      \hline
        0  & $5.07$ & $92.03$  & $127.97$ & $2.49\times10^{-2}$ & $4.43\times10^{-4}$ \\ \hline
        1  & $5.02$ & $47.59$  & $77.22$  & $1.93\times10^{-2}$ & $3.60\times10^{-4}$ \\ \hline
        2  & $4.82$ & $67.94$  & $92.29$  & $1.78\times10^{-2}$ & $3.75\times10^{-4}$ \\ \hline
        3  & $4.89$ & $70.03$  & $59.18$  & $1.28\times10^{-2}$ & $1.83\times10^{-3}$ \\ \hline
        4  & $5.09$ & $82.70$  & $83.71$  & $2.05\times10^{-2}$ & $5.40\times10^{-4}$ \\ \hline
        5  & $4.80$ & $124.04$ & $29.51$  & $2.11\times10^{-1}$ & $4.08\times10^{-4}$ \\ \hline
        6  & $5.20$ & $42.87$  & $59.87$  & $1.09\times10^{-2}$ & $4.34\times10^{-4}$ \\ \hline
        7  & $5.14$ & $95.38$  & $45.33$  & $2.20\times10^{-2}$ & $4.56\times10^{-4}$ \\ \hline
        8  & $5.07$ & $100.04$ & $88.11$  & $1.60\times10^{-2}$ & $3.36\times10^{-4}$ \\ \hline
        9  & $5.17$ & $102.78$ & $126.37$ & $1.16\times10^{-2}$ & $5.31\times10^{-4}$ \\ \hline
        10 & $4.92$ & $67.52$  & $49.84$  & $1.21\times10^{-2}$ & $3.94\times10^{-4}$ \\ \hline
        11 & $4.96$ & $85.88$  & $92.96$  & $2.42\times10^{-2}$ & $2.83\times10^{-4}$ \\ \hline
        12 & $5.04$ & $17.98$  & $78.72$  & $2.96\times10^{-2}$ & $8.17\times10^{-4}$ \\ \hline
        13 & $5.11$ & $69.75$  & $151.39$ & $1.04\times10^{-2}$ & $5.31\times10^{-4}$ \\ \hline
        14 & $4.90$ & $46.68$  & $71.60$  & $1.04\times10^{-2}$ & $4.23\times10^{-4}$ \\ \hline
        15 & $4.85$ & $43.04$  & $59.59$  & $1.03\times10^{-1}$ & $4.45\times10^{-4}$ \\ \hline
        16 & $5.02$ & $103.24$ & $59.52$  & $4.80\times10^{-2}$ & $1.35\times10^{-3}$ \\ \hline
        17 & $5.05$ & $50.87$  & $96.10$  & $3.32\times10^{-2}$ & $2.26\times10^{-3}$ \\ \hline
        18 & $4.94$ & $81.45$  & $86.48$  & $1.72\times10^{-2}$ & $4.92\times10^{-4}$ \\ \hline
        19 & $4.76$ & $72.08$  & $62.28$  & $9.20\times10^{-3}$ & $5.40\times10^{-4}$ \\ \hline
        20 & $5.04$ & $92.45$  & $66.36$  & $1.89\times10^{-2}$ & $3.64\times10^{-4}$ \\ \hline
        21 & $4.83$ & $88.85$  & $36.02$  & $5.79\times10^{-2}$ & $4.61\times10^{-4}$ \\ \hline
        22 & $4.99$ & $96.86$  & $45.13$  & $1.30\times10^{-2}$ & $2.93\times10^{-4}$ \\ \hline
        23 & $5.13$ & $81.18$  & $89.89$  & $1.47\times10^{-2}$ & $2.90\times10^{-4}$ \\ \hline
        24 & $4.98$ & $113.32$ & $137.50$ & $1.15\times10^{-2}$ & $3.70\times10^{-4}$ \\ \hline
        25 & $4.85$ & $109.80$ & $68.91$  & $3.34\times10^{-2}$ & $2.73\times10^{-4}$ \\ \hline
        26 & $4.96$ & $108.41$ & $113.01$ & $6.90\times10^{-3}$ & $3.25\times10^{-4}$ \\ \hline
    \end{tabular}
\end{table*}

\FloatBarrier
\bibliographystyle{apsrev4-2-mod}
\bibliography{reference.bib}
\end{document}